\documentclass[a4paper,11pt]{article}
\usepackage{jheppub} 
\usepackage{amsmath}
\usepackage{enumitem}

\newtheorem{mydef}{Definition}
\newtheorem{mythe}{Theorem}
\newtheorem{my}{Conjecture}

\def\beq#1\eeq{\begin{align}#1\end{align}}

\title{ Three dimensional canonical singularity and five dimensional $\mathcal{N}=1$ SCFT }

\author[b,c]{Dan Xie}
\author[a,b,c]{Shing-Tung Yau}

\affiliation[a]{Department of Mathematics, Harvard University, Cambridge, MA 02138, USA}
\affiliation[b]{Center of Mathematical Sciences and Applications, Harvard University, Cambridge, 02138, USA}
\affiliation[c]{Jefferson Physical Laboratory, Harvard University, Cambridge, MA 02138, USA}

\abstract{We conjecture that every three dimensional canonical singularity defines a five dimensional $\mathcal{N}=1$ 
SCFT. Flavor symmetry can be found from singularity structure: non-abelian flavor symmetry is read from the singularity type over one dimensional singular locus. The dimension of Coulomb branch is given by the number of compact crepant divisors from a crepant resolution of singularity. The detailed structure of Coulomb branch is described as follows:
a): A chamber of Coulomb branch is described by a crepant resolution, and this chamber is given by  its Nef cone and  the prepotential is computed from triple intersection numbers; b): Crepant resolution is not unique and different resolutions are related by flops; Nef cones from crepant resolutions form a fan which is claimed to be the full Coulomb branch. }

\begin{document} 
\maketitle
\flushbottom

\section{Introduction}
One can define higher dimensional superconformal field theory (SCFT) in various ways. If our theory has a conformal manifold, it might be possible to find a weakly coupled gauge theory description: one can describe our theory by specifying matter 
contents and gauge groups, and the coordinates of conformal manifold are identified with gauge couplings. 
This method works for four dimensional $\mathcal{N}=4$ Super Yang-Mills theory 
and four dimensional $\mathcal{N}=2$ SCFTs. These gauge theory descriptions are often not unique and one can have very interesting 
$S$ duality property.

Another way of defining a SCFT is as follows \cite{Xie:2015rpa}: consider the Coulomb branch of a four dimensional $\mathcal{N}=2$ theory, and there might be
a locus where some extra massive BPS particles become massless, and one can get a SCFT if these extra massless particles are mutually non-local. Argyres-Douglas theory \cite{Argyres:1995jj}
was found in this way, and a very general argument for the existence of such SCFT is given in \cite{Argyres:1995xn}. These extra massless degree of freedoms make
the description of physics singular at this particular point, and such singularities can often be made geometrical \cite{Gukov:1999ya}.  One can actually define a SCFT
by simply specifying the geometrical singularity, and this idea has been used  to 
engineer a large class of new four dimensional $\mathcal{N}=2$ SCFTs \cite{Xie:2015rpa,Chen:2016bzh,Wang:2016yha} . 
  
Our focus in this paper is to use singularity approach to study five dimensional $\mathcal{N}=1$ SCFT \cite{Seiberg:1996bd}. This type of SCFT has no SUSY preserving exact marginal deformations \cite{Cordova:2016xhm}, therefore one 
can not write down a weakly coupled gauge theory description with conformal gauging. 
One can still define 5d  $\mathcal{N}=1$ SCFT \cite{Seiberg:1996bd} by going to a locus of Coulomb branch where one has extra massless particles including  W bosons, instanton particles and tensionless strings \footnote{We want to emphasize that it is 
not sufficient to define a 5d $\mathcal{N}=1$ SCFT by just specifying  non-abelian gauge theory description  on certain locus of Coulomb branch, and one need to also specify other massive degree frames such as
instanton particles and strings which become massless or tensionless at the SCFT point.}. 
Again, instead of specifying extra massless degree of freedoms, we may use a geometric singularity to define a 5d $\mathcal{N}=1$ SCFT. Such method 
has been used in \cite{Morrison:1996xf,Intriligator:1997pq} to study 5d $\mathcal{N}=1$ SCFT and the main purpose of this paper is to provide a more systematic treatment. 
 
The basic idea of engineering 5d $\mathcal{N}=1$ SCFT is using M theory on a 3-fold singularity \cite{Morrison:1996xf,Intriligator:1997pq}. 
Then the first question that we would like to answer is that what kind of singularity will lead to a five dimensional $\mathcal{N}=1$ SCFT?  The main conjecture of this paper is 
\begin{my}
M theory on a 3d canonical singularity $X$ defines a 5d $\mathcal{N}=1$ SCFT.
\end{my}
Familiar examples of 3d canonical singularity include toric Gorenstein singularity, quotient singularity $C^3/G$ with $G$ a finite subgroup of $SL(3)$, and certain class of hypersurface singularities. 
One of basic argument for this conjecture is that this is the class of singularities that would appear in the degeneration limit of Calabi-Yau manifold \cite{wang1997incompleteness}.
Two dimensional canonical singularity has a ADE classification, and this leads to a remarkable ADE classification of six dimensional $(2,0)$ SCFT \cite{Witten:1995zh}.  

The two very basic numeric invariants associated with a 5d SCFT are the rank  of flavor symmetry $f$, and the rank  of the Coulomb branch  $r$. Finer information such 
as the enhancement of flavor symmetry, chamber structure  and  prepotential of Coulomb branch are also desired. 
One can derive many of these important physical results from the following geometric properties of singularity $X$:
\begin{itemize}
\item The non-abelian flavor symmetry can be read from the ADE type over one dimensional singular locus.
The rank of other abelian flavor symmetry is read from the rank of local divisor class group of the singularity. 

\item The Coulomb branch is given by  \textbf{crepant} resolution of the singularity $X$, and different crepant resolutions describe different chambers of the Coulomb branch. 
These different resolutions are related by flops, and the rank of Coulomb branch is constant across different chambers.
\item Given a crepant resolution, let's choose a basis $D_i$ of generators of local divisor class group, and the preopotential is given by  following formula \cite{Intriligator:1997pq}:
\begin{equation}
{\cal F}={1\over 6} (\sum_i \phi_i D_i)^3={1\over 6}\sum (D_i\cdot D_j \cdot D_k) \phi_i \phi_j \phi_k.
\end{equation}
Here $(D_i\cdot D_j \cdot D_k)$ is the triple intersection number for divisors. If $D_i$ is compact (resp. non-compact), the corresponding parameter $\psi_i$ is regarded as Coulomb branch (mass) parameter.  
The range of real numbers $\psi$ is determined by Nef cone. Inside Nef cone, one have an abelian gauge theory description in the IR.  The co-dimensional one face of 
Nef cone describes the place where an extra massive particle becomes massless. At the intersection of these faces, more massive particles become massless, and sometimes one can have a non-abelian gauge theory description.  

\item For crepant resolution related by a flop, the corresponding Nef cones share a face. Nef cones from all crepant resolutions form a fan which is identified as the full Coulomb branch.

\end{itemize}

This paper is organized as follows: section 2 reviews some basic facts of 5d $\mathcal{N}=1$ SCFT and 5d gauge theory; Section 3 describes the classification of 3-fold canonical singularity. 
Section 4 discusses in detail the SCFT associated with toric canonical singularity; Section 5 and 6 describe  SCFTs associated with quotient  and hypersurface singularity; 
Finally a conclusion is given in section 7.

\section{Generality of 5d $\mathcal{N}=1$ SCFT}
5d $\mathcal{N}=1$ superconformal algera consists of conformal group $SO(2,5)$, $R$ symmetry group $ SU(2)_R$, and  possibly
global symmetry group $G$. Representation theory of five dimensional $\mathcal{N}=1$ superconformal algebra is studied in \cite{Cordova:2016xhm,Buican:2016hpb}, and a general multiplet is labeled as $[j_1,j_2]_\Delta^{(R)}$, here $j_1, j_2, \Delta$ are
spins and scaling dimension, and $R$ is $SU(2)_R$ quantum number. Short representations are classified in \cite{Cordova:2016xhm,Buican:2016hpb}, and one important class is called 
$C_R=[0,0]_{\Delta}^{(R)}$, with the relation $\Delta={3\over2}R$. Flavor currents is contained in multiplet $C_1$. 

For 5d $\mathcal{N}=1$ SCFT, the only SUSY preserving relevant deformation is the mass deformation, and there is \textbf{no} exact marginal deformation. 
One can have a different kind of SUSY preserving deformation by turning on expectation value of certain operators, and we could have continuous space of vacua called moduli space. The moduli space 
consists of various branches:
\begin{itemize}
\item There could be a Higgs branch which can be parameterized by the expectation value of operators $C_R$,  
and the $SU(2)_R$ symmetry acts non-trivially on this branch. We would like to determine the full chiral ring relation of the Higgs branch, from which we can 
read the non-abelian flavor symmetry $G$. The simplest data we'd like to determine is 
the rank $f$ of the flavor group $G$. 
\item There could be a Coulomb branch which is parameterized by the real numbers $\psi_j,~j=1,\ldots r$. 
Unlike four dimensional $\mathcal{N}=2$ theory, we can not parameterize it by expectation value of BPS operators.
The Coulomb branch is not lifted by turning on mass deformations with mass parameters $m_i,~i=1,\ldots,f$, but the low energy physics is changed.
At generic point, the low energy theory could be described by abelian gauge theory with gauge group $U(1)^r$, and the important question is to determine the rank $r$ and 
the prepotential which is a cubic function of  $U(1)$ vector multiplet and mass parameter: 
\begin{equation}
{\cal F}(m,\psi)= \sum_{i,j,k}a_{ijk}\phi_i\phi_j\phi_k.
\end{equation} 
We have $\phi_{i}=m_i,~i=1,\ldots, f$, and $\phi_{f+j}=\psi_j,~~j=1,\ldots,r$.  A particular intriguing feature of 5d $\mathcal{N}=1$ SCFT is that there might be many chambers of Coulomb branch and 
each chamber has different prepotentials.  The BPS particle has central charge $Z=n_e^{(i)} \psi_i+S_i m_i$, and one also has tensile strings whose tension is given by $Z= n_m^{(i)} \psi_i^D$ with $\psi_i^D$  the dual coordinates.  
\item One could also have the  mixed branch which is a direct product of a Coulomb and Higgs factor. 
\end{itemize}

Five dimensional $\mathcal{N}=1$ SCFT can be constructed as follows:
\begin{enumerate}
\item Particular UV completion of a non-abelian gauge theory \cite{Seiberg:1996bd,Morrison:1996xf}, and the typical example is $SU(2)$ with $N_f\leq 7$ fundamental flavors and the corresponding UV SCFT is  5d $E_{N_f+1}$ SCFT. 
\item M theory on 3-fold singularities \cite{Witten:1996qb,Morrison:1996xf,Intriligator:1997pq, Douglas:1996xp, Leung:1997tw, Katz:1997eq}, and the cone over Del Pezzo $k$ surfaces will give 5d $E_{k}$ SCFT.
\item $(p,q)$ five brane webs in type IIB string theory \cite{Aharony:1997bh}, and some of  $E_{k}$ SCFTs can be engineered in this way.
\end{enumerate}
Many aspects of 5d $\mathcal{N}=1$ SCFTs  such as the verification of enhanced flavor symmetry, superconformal inex, etc have been recently studied in \cite{Iqbal:2012xm,Kim:2012qf,Assel:2012nf,Rodriguez-Gomez:2013dpa,Bergman:2013koa,Kim:2013nva,Bao:2013pwa,Bergman:2013ala,Hayashi:2013qwa,Taki:2013vka,Bergman:2013aca,Taki:2014pba,Hwang:2014uwa, Zafrir:2014ywa,Esole:2014bka, Esole:2014hya,Hayashi:2014wfa, Bergman:2014kza,Mitev:2014jza,Gaiotto:2014ina,Lambert:2014jna,Tachikawa:2015mha, Zafrir:2015uaa, Yonekura:2015ksa, Hayashi:2015xla, Kim:2015jba, Zafrir:2015ftn, Bergman:2015dpa, Gaiotto:2015una, Bergman:2016avc, Qiu:2016dyj, Chang:2016iji, DHoker:2016ysh, DHoker:2017mds, Esole:2017kyr, Hayashi:2017jze}.

\subsection{5d Gauge theory and enhanced flavor symmetry}
Along some sub-locus of Coulomb branch, the low energy theory can be described by a non-abelian gauge theory.  
Part of enhanced flavor symmetry might be derived by analyzing the instanton operators \cite{Lambert:2014jna,Tachikawa:2015mha,Zafrir:2015uaa,Yonekura:2015ksa,Bergman:2016avc}, and here we review the basic result. 
In this subsection, all the gauge groups are taken as $SU(N)$ and we only consider bi-fundamental matter. Consider a single $SU(N)$ gauge group coupled with $n_f$ hypermultiplets in fundamental representation, one can turn on 
Chern-Simons term with level $k$ satisfying the constraint 
\begin{equation}
k+{n_f\over 2} \in Z.
\end{equation}
We use $m_0={1\over g_{cl}^2}$ to denote the classical gauge coupling. Part of coulomb branch can be parameterized by the real numbers $(a_1,\ldots, a_N)$ with the constraint $\sum a_i=0$. Using Weyl invariance, 
we can take $a_1\geq a_2\geq \ldots \geq a_N$. At $m_0=0$, the prepotential reads
\begin{equation}
{\cal F}={1\over 6}(\sum_{i<j}^N(a_i-a_j)^3+ k\sum_{i=1}^3 a_i^3-{n_f\over2}\sum_{i=1}^N|a_i|^3).
\end{equation}
The manifest flavor symmetry is $U(n_f)\times U(1)_I$, where $U(1)_I$ is associated with the instanton current $*F\wedge F$. The SCFT point might be achieved by taking $a_i$, gauge couplings, and masses to be zero \footnote{To completely specify the SCFT, we need to specify other BPS particles which become massless at SCFT point besides the W bosons.}, and the flavor symmetry 
at the SCFT point could be further enhanced. 

We will also consider a linear quiver with $SU(N_i)$ gauge group $[n_i]-SU(N_1)-SU(N_2)-\ldots-SU(N_r)-[n_e]$. We also have the Chern-Simons level $k_i$ on each gauge group $SU(N_i)$, and we put further constraints on $k_i$: 
\begin{equation}
2|k_i|\leq 2N_i-N_f,
\end{equation}
here $N_f$ is the number of flavors on  $i$th gauge group. The manifest flavor symmetry is $U(n_i)\times U(n_e) \times U(1)^{r}\times U(1)^{r-1}$, with $U(n_i)$ and $U(n_e)$ from the fundamental flavors on two ends of quiver, and $U(1)$s from 
instanton current and flavor of bi-fundamentals.  The enhanced flavor symmetry can be found as follows. Consider two $A_r$ Dynkin diagrams $\Gamma^{\pm}$, and for $\Gamma^{+}$, we color $i$th node of $\Gamma^{+}$ black 
if the following identity holds
\begin{equation}
2k_i=2N_i-N_f.
\end{equation}
Similarly, color $i$th node of $\Gamma^{-}$ black if we have  
\begin{equation}
-2k_i=2N_i-N_f.
\end{equation}
The flavor symmetry from $\Gamma^{+}$ is $G_{\Gamma^{+}}=\bigotimes_j A_{n_j} \times U(1)^{r-\sum n_i}$, here $n_j$ is the length of $j$th connected sub-diagram whose vertices are black. Similarly, one can read the enhanced flavor symmetry $G_{\Gamma^{-}}$ from $\Gamma^{-}$. 
If we have the quiver tail $2-SU(2)-SU(3)-\ldots$, we replace the quiver as $SU(1)-SU(2)-SU(3)-\ldots$ and apply the above rule again. 
The overall flavor symmetry for the linear quiver is 
\begin{equation}
G_f=G_{\Gamma^{+}}\times G_{\Gamma^{-}}\times U(n_i)\times U(n_e)/U(1). 
\end{equation}
The flavor symmetry might be further enhanced.

\section{Canonical singularity and five dimensional $\mathcal{N}=1$ SCFT}

\subsection{Three dimensional canonical singularity}
We conjectured in the introduction that 3-fold canonical singularities would lead to 5d $\mathcal{N}=1$ SCFT. In this section, we will review the relevant facts about 
3-fold canonical singularity \cite{reid1979canonical,reid1983minimal,reid1985young}. Let's start with the definition:

\begin{mydef}
A variety $X$ is said to have canonical singularity if its is normal and the following two conditions are satisfied
\begin{itemize}
\item the Weil divisor $K_X$ \footnote{$K_X$ is the canonical divisor associated with X.} is Q-Cartier, i.e. $rK_X$ is a Cartier divisor \footnote{A Cartier divisor implies that it can be used to define a line bundle.}. Here $r$ is called index of the singularity.
\item for any resolution
\footnote{ Given an irreducible variety $X$, a resolution of singularities of $X$ is a morphism $f: X^{'}\rightarrow X$  such that:
\begin{enumerate}[label=\alph*]
\item: $X^{'}$ is smooth and irreducible.
\item: $f$ is proper.
\item: $f$ induces an isomorphism of varieties $f^{-1}(X / X_{sing}) = X /X_{sing}$. Furthermore, $f : X^{'}\rightarrow X$ is a projective resolution if $f$ is a projective morphism.
\end{enumerate}}
of singularities $f: Y\rightarrow X$, with exceptional divisors $E_i\in Y$, the rational numbers $a_i$ satisfying 
\begin{equation}
K_Y=f^{*} K_X+\sum a_i E_i.
\end{equation}
are nonnegative. The numbers $a_i$ are called the discrepancies of $f$ at $E_i$; if they are all positive $X$ is said to have  \textbf{terminal} singularities. If $a_i=0$, the corresponding prime exceptional divisor is 
called crepant divisor.
\end{itemize}
\end{mydef}

Two dimensional canonical singularity is completely classified and is also called Du Val singularity. There is a ADE classification, and it can be characterized by the following three ways:
\begin{itemize}
\item They are described by the following hypersurface singularity:
\begin{align}
&A_{n}:~~x_1^2+x_2^2+x_3^{n+1}=0, \nonumber\\
&D_{n}:~~x_1^2+x_2^{n-1}+x_2x_3^{2}=0, \nonumber\\
&E_{6}:~~x_1^2+x_2^3+x_3^{4}=0, \nonumber\\
&E_{7}:~~x_1^2+x_2^3+x_2x_3^{3}=0, \nonumber\\
&E_{8}:~~x_1^2+x_2^3+x_3^{5}=0.
\end{align}
\item They are defined as the quotient singularity $C^2/G$ with $G$ a finite subgroup of $SL(2)$.
\item For $A$ type canonical singularity, one has a toric description. 
\end{itemize}
Let's first introduce the concept of crepant resolution. 
Let X be a variety with canonical singularities. A partial resolution of X is a proper birational morphism $\phi: Y\rightarrow X$ from a normal variety. The morphism $\phi$ is said to be \textbf{crepant} if 
\begin{equation}
K_Y=\phi^{*} K_X.
\end{equation}
Two dimensional canonical singularity actually has a crepant resolution where $Y$ is smooth. 

Three dimensional canonical singularity is not completely classified, but a lot of general properties have been studied by Reid \cite{reid1979canonical,reid1983minimal,reid1985young}. There are some useful properties for 3-fold canonical singularities:
\begin{enumerate}
\item Canonical singularity implies Du Val singularities in codimension 2. This implies that the singularity type over a one dimensional singular locus has to be of ADE type. 
\item  Every canonical singularity can be 
written as $X={\tilde{X}/ \mu_r}$  with $\tilde{X}$ an index one canonical singularity. 
The index one canonical singularity is also called rational Gorenstein singularity. 
\item There exists partial crepant resolution of a canonical singularity:

\begin{mythe}
Let X be an algebraic 3-fold with canonical singularities. Then there exists a crepant projective morphism $\phi: Y\rightarrow X$, which is an isomorphism in codimension 1, from a 3-fold Y with Q-factorial terminal singularities. 
\end{mythe}
A Q-factorial terminal singularity means that every Weil divisor is a Q-Cartier divisor. Such \textbf{crepant} resolution is not unique, and different resolutions are related by \textbf{flop}. The crucial fact is that the 
number of crepant divisors are independent of the choice of crepant resolution. In the following, by crepant resolution we always mean  above crepant morphism. 

\end{enumerate}

\subsubsection{Classification of rational Gorenstein singularity}
Since an index $r$ canonical singularity can be covered using an index one canonical singularity, we will focus on index one canonical singularity which is 
also called rational Gorenstein singularity. To a rational Gorenstein 3-fold singularity, the general hyperplane section $H$ through the singularity is a two dimensional rational or elliptic Gorenstein singularity \cite{reid1979canonical}.
Two dimensional rational Gorenstein singularity is nothing but the ADE singularity, and elliptic Gorenstein singularity is next simplest surface singularity. 
The above two class of surface singularities are quite rigid which makes the classification much simpler. In fact, use the minimal resolution of surface singularity, 
one can attach a natural integer $k\geq 0$ for each 3-fold rational Gorenstein singuarlity.  We have \cite{reid1979canonical}:
\begin{itemize}
\item $k=0$, the singularity is a cDV point, i.e. the singularity can be written as follows 
\begin{equation}
f(x,y,z)+tg(x,y,z, t)=0,
\end{equation}
here $f(x,y,z)$ denotes the Du Val singularity. 
\end{itemize}
 $k\geq 1$, the general section $H$ through the singularity has an elliptic Gorenstein singularity $p\in H$ with invariant k:
\begin{itemize}
\item If $k\geq 2$, then $k=mult_P X$. 
\item If $k\geq 3$, then $k+1$ is equal to the embedding dimension $=\text{dim} (m_p/m_p^2)$.
if $k=2$, then $P\in X$ is isomorphic to a hypersurface given by $x^2+f(y,z,t)$ with $f$ a sum of monomials with degree bigger than 4. If $k=1$, then 
then $P\in X$ is isomorphic to a hypersurface given by $x^2+y^3+f(y,z,t)=0$ with $f(y,z,t)=yf_1(z,t)+f_2(z,t)$ and $f_1$ (respectively $f_2$) is a sum of 
monomials $z^a t^b$ of degree $\geq 4$ (respectively $\geq 6$).
\item If $k=3$, the singularity $P$ is given by a hypersurface.
\item If $k=4$, then $P$ is  given by a complete intersection defined by two polynomials $(f_1, f_2)$. 
\item For $k\geq 5$, the classification is incomplete. 
\end{itemize} 
Once we classify the rational Gorenstein singularity, the next step is to find out which cyclic quotient $\mu_r$ would lead to 
a general canonical singularity. 

Other class of familiar examples include the toric Gorenstein singularity which will be studied in next section, and the quotient singularity $C^3/G$ with $G$
a finite subgroup of $SL(3)$. 

\subsubsection{Classification of terminal singularity}
The 3-fold terminal singularity has been completely classified. Terminal singularity is important as they appear at the end of partial crepant resolution. 
 A 3-fold singularity is called cDV singularity if it can be written as the following form
\begin{equation}
f(x_1, x_2, x_3)+t g(x_1, x_2, x_3, t)=0.
\end{equation}
Here $f(x_1, x_2, x_3)$ is the two dimensional Du Val singularity.  
Index one terminal singularity is classified by \textbf{isolated cDV} singularity. 

Other terminal singularities are found as the cyclic quotient
of the isolated cDV singularity, and is  classified by Mori \cite{mori19853}. The major result is that terminal singularities belong to 6 families listed in table. \ref{table:terminal}. 

\begin{table}[h]
\begin{center}
\begin{tabular}{|c|c|c|c|c|}
\hline 
~& $r$ &Type&$f$&Conditions  \\ \hline
(1)& $any$ & ${1\over r}(a,-a,1,0;0)$& $xy+g(z^r,t)$ & $g\in m^2,~a,r,\text{coprime}$ \\ \hline
(2)&4& ${1\over4}(1,1,3,2;2)$ & $xy+z^2+g(t)$ & $g\in m^3$ \\ \hline
~&~&~&or $x^2+z^2+g(y,t)$& $g\in m^3$ \\ \hline
(3)&2 & ${1\over2}(0,1,1,1;0)$ & $x^2+y^2+g(z,t)$ & $g\in m^4$ \\ \hline
(4)& 3 & ${1\over3}(0,2,1,1;0)$ &$x^2+y^3+z^3+t^3$ &~ \\ \hline
~&~&~&or $ x^2+y^3+z^2t+yg(z,t)+h(z,t)$ & $g\in m^4,h\in m^6$ \\ \hline
(5)&2&${1\over 2}(1,0,1,1;0)$&or $x^2+y^3+y zt+g(z,t)$ & $g\in m^4$ \\ \hline
~&~&~&or $x^2+yzt+y^n+g(z,t)$& $g\in m^4,~n\geq 4$ \\ \hline
~&~&~&or $x^2+yz^2+y^n+g(z,t)$ & $g\in m^4,~n\geq 3$ \\ \hline
(6)&2&${1\over2}(1,0,1,1;0)$ & $x^2+y^3+yg(z,t)+h(z,t)$& $g,h\in m^4, h_4\neq 0$ \\ \hline
\end{tabular}
\end{center}
\caption{List of three-fold terminal singularity.}
  \label{table:terminal}
\end{table}
It is also proven in \cite{kollar1988threefolds} that every isolated singularity in above list is terminal. 
If we start with a smooth point (case (1) in table. \ref{table:terminal}), then the terminal singularity is classified as follows:
\begin{mythe}
X is terminal if and only if (up to permutations of $(x,y,z)$ and symmetries of $\mu_r$) $X=C^3/\mu_r$ of type ${1\over r}(a, -a, 1)$ with $a$ coprime to $r$, where $\mu_r$ is the cyclic 
group of order $r$.
\end{mythe}

\subsection{Physics and geometry}

\textbf{Flavor symmetry}: The flavor symmetry of  SCFT can be read from the singularity structure. 
The  non-abelian flavor 
symmetry can be read as follows: since our singularity is normal, the singular locus $\Gamma$ is at most one dimensional, 
and one has Du Val ADE surface singularity over one dimensional singular locus. 
Using the fact that one get ADE gauge symmetry by putting M theory on ADE singularity,
and since the singular locus is non-compact (affine), one get a corresponding ADE flavor group for each one dimensional component $\Gamma_i\subset \Gamma$. One also have abelian flavor symmetry 
associated with the local divisor class group, whose rank is denoted as $d$.
The full flavor symmetry is then
\begin{equation}
G=U(1)^{d}\prod_i G_i.
\end{equation}
Here $G_i$ is the corresponding $ADE$ flavor symmetry associated with 
$i$th component of singular locus. The flavor symmetry might be further enhanced.  We also use $f$ to denote the rank of G.

\textbf{Partial resolution and non-abelian gauge theory description}: Let's start with a canonical singularity, and  it might be possible to find a partial crepant resolution $f:Y\rightarrow X$ such 
that $Y$ has one dimensional singular \textbf{compact} locus whose singular type is one of ADE type. According to 
the similar M theory argument we used for finding flavor symmetry, we get ADE gauge symmetry. It is also possible to 
get a non-abelian gauge theory description with Lagrangian description. 

\textbf{Coulomb branch and crepant resolution}: Starting with  canonical singularity, one can find crepant partial resolutions $f:Y\rightarrow X$, here $Y$ has Q-factorial terminal
singularity.  For rational Gorenstein singularity, such partial resolution can be found by first using sequence of standard blow-up for $k\geq 3$. For $k=0$ or $k=1$, we use the following 
weighted blow up, and one affine piece of which is given by setting:
\begin{align}  
& k=2:~~x=z^2x_1,~y=zy_1,~z=z, \nonumber\\
&k=1:~x=z^3x_1,~y=z^2y_1,~z=z.
\end{align}
After these blow-ups, we end-up with a variety $Y$ with cDV singularity. We can further blow-up the one-dimensional Du Val singularity so that we 
get an isolated cDV singularity. One can further perform small resolution on non Q-factorial isolated cDV singularity, and the end result is that 
we get an variety Y with Q-factorial terminal singularity. Using the cyclic cover construction, one can construct such crepant  resolution for every canonical singularity. 

Such crepant resolutions are not unique, but these different resolutions are all related by a sequence flops!  Let's briefly introduce the concept of flops 
using the conifold singularity $X_0$: $x^2+y^2+z^2+w^2=0$. This singularity has two crepant resolutions $X_1$ and $X_2$ where the exceptional locus is a curve, 
we say that $X_1$ and $X_2$ are related by flop: first a curve $C_1$ in $X_1$ shrinks to zero size so we get singularity $X_0$, we then resolve $X_0$ in a different 
way to get another variety $X_2$ where the exceptional locus is another curve $C_2$. 

Let's now fix a crepant resolution $Y$, and one can define the Mori cone and Nef cone. A divisor $D$ is called Nef if 
\begin{equation}
D\cdot C \geq 0,
\end{equation}
for all complete curve $C$. All such Nef divisors form a cone and is called Nef-Cone. Its dual cone is called Mori cone, which can also 
be defined using the complete curve.  Our major conjecture is that the range of Coulomb branch is just the Nef Cone.

Let's denote the compact crepant divisors as $E_i,~~i=1,\ldots, r$, and non-compact crepant divisors as $D_i,~i=1,\ldots, f_1$, and further 
choose a basis  $D_j,~j=1,\ldots,f-f_1$ of non-compact divisors which generates the local class group of original singularity. 
The Coulomb branch deformation is related to the compact divisors $E_i$, while the mass deformation is related to $D_i$.
The prepotential takes the following form
\begin{equation}
F={1\over 6}(\sum_{i=1}^rE_i \phi_i+\sum_{n=1}^fD_n m_n)^3={1\over 6} \sum E_i\cdot E_j \cdot E_k \phi_i \phi_j \phi_k. 
\end{equation}
Here $E_i\cdot E_j \cdot E_k$ is the triple intersection form (Here we use $E$s to denote all the divisors).

One can do the similar computation for the Nef-cone and prepotential of different crepant partial resolutions. Since different crepant partial resolution 
is related by flop, and their Nef cones form a fan, which might be called Nef fan, our main conjecture in this paper is that 
\begin{my}
The Coulomb branch of 5d $\mathcal{N}=1$ SCFT is the Nef fan.
\end{my} 

The BPS spectrum can be described as follows: Let's start with the generators $C_i$ of Mori cone, and a M2 brane can wrap on those curves and form primitive BPS particles. On the other hand, M5 brane can wrap on the compact exceptional divisors and one get tensile strings. All these massive objects become massless in the SCFT limit. 

\section{Toric singularity}
An important and useful class of 3-fold canonical singularity can be defined using toric method \cite{cox2011toric}, and 
we call them toric singularity. Toric singularity can be defined in a combinatorial way and 
various of its properties such as crepant resolutions can be described explicitly, which makes many computations possible. We will describe these toric singularities which are also canonical singularity. 

We first briefly review how to define a toric variety, for more details, see \cite{cox2011toric}.  Let's start with a tree dimensional standard lattice $N$, and its dual lattice $M=\text{Hom}(N,Z)$. A rational convex polyhedral
 cone $\sigma$ in $N_R=N\otimes R$ is defined by a set of lattice vectors $\{ v_1,\ldots, v_n\}$ as follows:
\begin{equation}
\sigma=\{  r_1 v_1+r_2 v_2+\ldots+ r_n v_n | r_i\geq 0\}.
\end{equation}
Here ray generator $v_\rho$ of $\sigma$ is a lattice vector  such that it is not a multiple of another lattice vector. Its dual cone $\sigma^{\vee}$ is defined by the set in $M_R=M\otimes R$ such that 
\begin{equation}
\sigma^{\vee}=\{ w\cdot v\geq 0|w\in M_R, v\in \sigma \},
\end{equation}
here we used standard pairing between lattices $N$ and $M$.   From $\sigma^{\vee}$, one can define a semigroup $S=\sigma^{\vee}\cap M$, and the affine variety associated 
with $\sigma$ is 
\begin{equation}
X_\sigma=Spec(\sigma^{\vee}\cap M).
\end{equation}
We have the following further constraints on the cones:
\begin{itemize}
\item A strongly rational convex polyhedral cone (s.r.c.p.c) is a convex cone $\sigma$ such that $\sigma\cap (-\sigma)=0$. 
\item A simplicial s.r.c.p.c is a cone whose generators form a $R$ basis of $N_R$.  
\end{itemize}

A fan $\Sigma$ with respect to a lattice $N$ is a finite collection of of rational convex polyhedral
 cones in $N_R$ such that 
 \begin{itemize}
 \item any face $\tau$ of $\sigma$ belongs to $\Sigma$.
 \item for $\sigma_1, \sigma_2 \in \Sigma$, the intersection $\sigma_1\cap \sigma_2$ is a face of both $\sigma_1$ and $\sigma_2$.  
 \end{itemize}
 A fan is called simplicial if all the cones are simplicial. 
 By $|\Sigma|:=\cup \{ \sigma |\sigma\in \Sigma\}$ one denotes the support and $\Sigma(i)$ the set of all $i$ dimensional 
 cones of fan $\Sigma$. 

There is an important cone-orbit correspondence for toric variety: for each $k$ dimensional cone, one can associate a 
$3-k$ dimensional toric invariant orbit in $X_\sigma$. In particular, the toric divisors are determined by one dimensional cone, 
which is also specified by the ray generator $v_\rho$.

We are interested in affine toric singularity which is defined by a single cone $\sigma$. 
The singular locus of affine toric singularity is given by the sub-cone $\sigma_i$ whose generators do not form a Z-basis, and 
the singular locus corresponds to the orbit which is determined by $\sigma_i$.  For 3-fold 
toric singularity, the maximal dimension of singular locus is one dimensional which is then formed by a two dimensional sub-cone $\sigma_i$ in 
$\sigma$.  

A toric singularity is Q-Gorenstein if one can find a vector $m_\sigma$ in $M_Q$ \footnote{$M_Q$ denotes points in $M_R$ with rational coordinates.} such that $\langle m_\sigma, v_\rho\rangle=1$ for 
all the one dimensional generator $v_\rho$ of $\sigma$.  A toric singularity is Gorenstein if one can find a vector $H$ in $M$ such that $\langle H, v_\rho\rangle=1$ for all $\rho$.  Equivalently, one can choose a hyperplane in $N$ such that all the ray generators $\sigma_\rho$
lie on it.   For a 3-fold Q-Gorenstein singularity with 
index $r$, we take the hyperplane as $z=r$, and the ray generators form a two dimensional convex polygon $P$. Our 3-fold Q-Gorenstein 
toric singularity is then specified by a index $r$ and a two dimensional  convex polygon $P$ at $z=r$ plane.  

Let's start with a 3-fold Q-Gorenstein toric singularity. The classification of 
3-fold canonical and terminal singularity is given as follows:
\begin{itemize}
\item The cone $\sigma$ of 3-fold \textbf{canonical} singularity has the following property: there is no lattice points in $\sigma$ between the origin and the 
polygon $P$. In particular,  toric Gorenstein singularity is canonical. 

\item The cone $\sigma$ of 3-fold \textbf{terminal} singularity is characterized as follows: there is no lattice points in $\sigma$ between the origin and the 
polygon $P$, and furthermore there is no internal lattice points for $P$. 

\end{itemize}
 
Among toric terminal singularities, the Q-factorial terminal singularity is specified by smooth point and the quotient singularity of the type $C^3/\mu_r$, with 
$\mu_r={1\over r}(a,-a,1)$, here $r$ and $a$ are co-prime. Smooth point is the only $Q$ factorial Gorenstein terminal singularity. There is only one kind of non Q-factorial Gorenstein terminal singularity:  the conifold singularity which is defined by the equation: $x_1^2+x_2^2+x_3^2+x_4^2=0$.

\subsection{Toric Gorenstein singularity}
We focus on toric Gorenstein 3-fold singularity which is specified by a convex polygon $P$ (It is often called toric diagram.).  The only terminal singularity is the smooth point whose 
associated convex polygon is a triangle with no internal lattice points and no boundary lattice points.  We denote the number of lattice points on the boundary of $P$ as $N$, and the number of internal lattice points as $I$.  

\subsubsection{Flavor symmetry}
The enhancement of flavor symmetry is determined by the ADE type of one dimensional singular locus, which in the toric case,
is associated with the two dimensional cone $\sigma_i$ in $\sigma$.  The two dimensional cone in $\sigma$ is specified by boundary edges of $P$.
For a one dimensional boundary in $P$, it is easy to determine the singularity type: 
the only toric ADE singularity is $A$ type, and the corresponding singularity type for  a boundary edge with $d$ internal points is $A_{d}$, and the flavor symmetry has the following type
\begin{equation}
\prod_i A_{d_i} \times U(1)^{N-3-\sum d_i}.
\end{equation}
The local divisor class group has rank $N-3-\sum d_i$, which is equal to the number of vertices of $P$ minus three.
So the rank of the mass parameter is given by the following formula
\begin{equation}
f=N-3.
\end{equation}

\begin{figure}[h]
    \centering
    \includegraphics[width=2.0in]{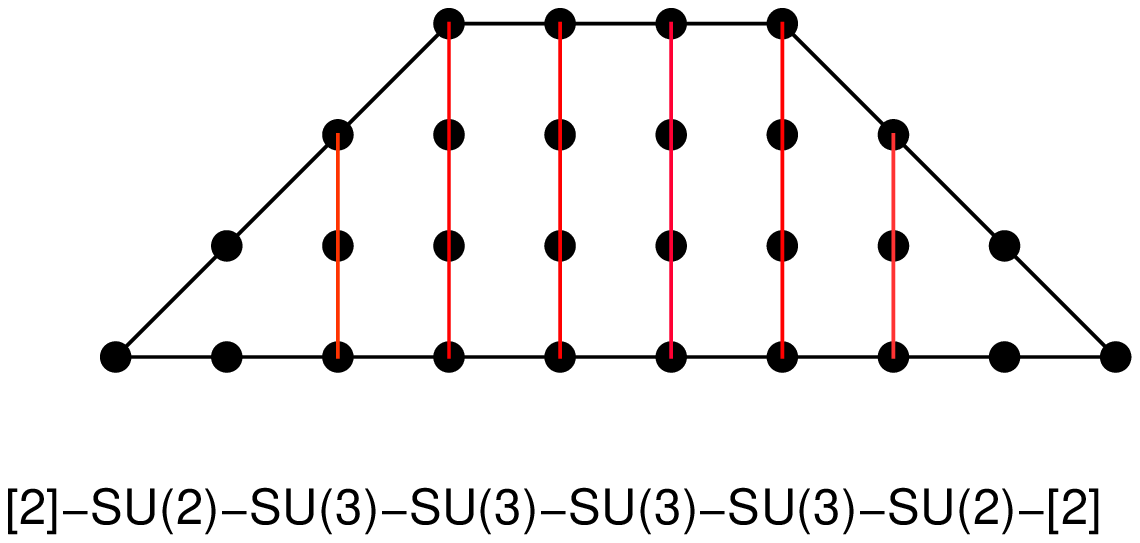}
    \caption{A toric diagram and one of its partial resolution, and we also write down a gauge theory description.}
    \label{flavor}
\end{figure}

\textbf{Example}: Let's see figure. \ref{flavor} for a toric diagram, and the flavor symmetry is 
\begin{equation}
G= SU(9)\times SU(3) \times SU(3) \times SU(3) \times U(1)
\end{equation}
We also write down a gauge theory description for it. Using the instanton method reviewed in section II (see \cite{Yonekura:2015ksa} and use the fact that the two $SU(3)$ gauge groups connected with $SU(2)$ gauge groups have CS level $k={1\over2}$.), we see that 
field theory result and geometric result agree.

\subsubsection{Partial resolution and gauge theory description}
One can consider partial resolution which is formed by adding edges connecting the vertices of $P$. 
Those internal edges represent non-abelian gauge group of $A_{d_i}$ type if 
there are $d_i$ internal points on extra edges. Those internal edges will cut $P$ into various smaller convex 
polygons $P_i$.  The physical interpretation is following: one have a non-abelian gauge theory description: each $P_i$ represents 
a matter system, and each added internal edge connecting two smaller polygons $P_A,P_B$ represents a gauge group  coupled to
two matter systems represented by $P_A$ and $P_B$. Sometimes we also have non-trivial CS term too.

For some convex polygon $P$, it is possible to find above type of subdivision of $P$ such that each subsystem $P_i$ carries no internal 
lattice points. Those $P_i$ could be interpreted as free matter, and one would have a non-abelian Lagrangian description. 

Often one could find more than one type of such decompositions of $P$, and one have different Lagrangian descriptions. It might be tempting to call 
these different descriptions as some kind of duality. However, we would like to avoid that because the description of 
non-abelian gauge theory itself does not define a SCFT.

\subsubsection{Crepant resolution and Coulomb branch solution}
\begin{figure}[h]
    \centering
    \includegraphics[width=2.0in]{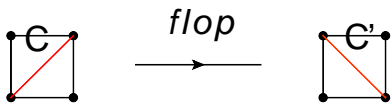}
    \caption{Two different resolutions of a lattice polygon can be related by sequence of flops shown above.}
    \label{flop}
\end{figure}

The  resolution of toric  singularity can be described in following explicit way: one construct a new fan $\Sigma$ from the cone $\sigma$ by adding 
rays inside $\sigma$. The \textbf{crepant} resolution  is given by the uni-modular triangulation of the lattice polygon $P$. Such triangulation 
is easily found by using the lattice points in polygon $P$ \footnote{For a two dimensional lattice polygon, its area is given by the formula $A={B\over 2}+I-1$, here $B$ is the boundary lattice points, 
and $I$ is the internal points. Uni-modular triangulation implies that all the triangles in the triangulation has area ${1\over 2}$. This is achieved by a triangle with no internal lattice points and no other boundary points except three vertices.
}.  The final toric variety is smooth as the only Q-factorial toric Gorenstein singularity is smooth. Obviously, such triangulations are not unique, but 
they can all be related by the \textbf{flops}, see figure. \ref{flop}. 

\textbf{Exceptional divisors} Let's choose a crepant triangulation $P^{'}$ of $P$. The exceptional divisors corresponding to the internal 
and boundary lattice points of $P$ which are not vertices. The number of such divisors is a constant  and is independent of choice of crepant resolution.  
The local divisor class group of $P^{'}$ is generated by the ray generators $v_\rho$ subject to following relations 
\begin{equation}
\sum_{\rho} \langle v_\rho, e_i \rangle D_\rho=0,~~i=1,2,3.
\end{equation}
Here $D_\rho$ is the corresponding divisor from $v_\rho$, and
 $e_i$ is the standard basis of $M$ and we used the bracket to denote the standard paring between $N$ and $M$. So there are three equations and the number of independent divisors are $N+I-3$.  The above relation is independent of the triangulation and only depend on polygon $P$. The topology of these divisors are follows: a): The divisors  from internal lattice points are compact; b): The divisors from vertices of the lattice polygon are non-compact, i.e. of the form $C^2$; c): The divisors from other boundary lattice points  are semi-compact, i.e. of the form $C\times P^1$. 

\begin{figure}[h]
    \centering
    \includegraphics[width=3.5in]{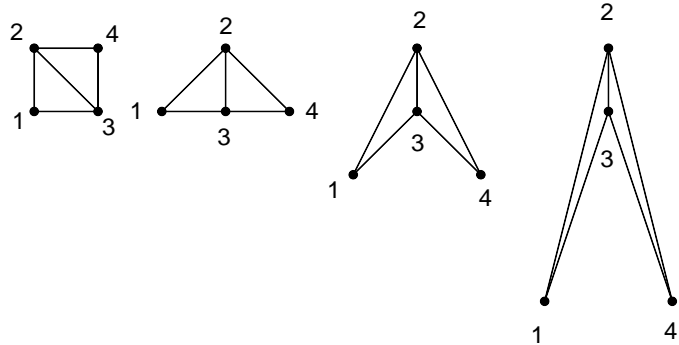}
    \caption{Quadrilaterals in a triangulation of a polygon $P$.}
    \label{quodri}
\end{figure}

\textbf{Triple intersection number}: Consider a quadrilateral in the triangulation whose vertices have coordinates $u_i,~i=1,\ldots,4$. We assume that the diagonal $C$ connects $u_2$ and $u_3$. The coordinates $u_i$ of  quadrilateral's vertices satisfy the relation:
\begin{equation}
u_1+a u_2 +b u_3 + u_4=0.
\end{equation} 
Since the $z$ coordinate of $u_i$s are all 1,  we have $a+b=-2$. The only non-trivial triple intersection numbers involving the curve $C=E_2\cdot E_3$ are 
\begin{equation}
C\cdot E_1 =1,~C\cdot E_4 =1,~C\cdot E_2 =a,~C\cdot E_3 =b.
\end{equation}
We have following solutions (see figure. \ref{quodri}): 
\begin{align}
& \text{Case 1}:~~a=-1,~b= -1 \nonumber\\
& \text{Case 2}:~~a=0,~~~b= -2 \nonumber\\
& \text{Case 3}:~~a=1,~~~b=-3 \nonumber\\
&\ldots \ldots \ldots \nonumber\\
& \text{Case i}:~~~a=i-2,~~b=-i
\end{align}
Notice that only case 1 allows the flop.  Using above formula, we can compute all the triple intersection number involving at least one compact divisors $E$. 
The  self-intersection number for a compact divisor $E$ can be computed as follows: we always have a relation between divisors
\begin{equation}
E+\sum D_i =0~\rightarrow~E^3=-\sum E^2 D_i,
\end{equation}
and one can get $E^3$ using the known triple intersection number. In fact only the divisors connected to 
 $E$ contribute to $E^3$.  Other intersection numbers involving no compact divisors can be computed using three relations between the divisors and above known intersection numbers involving at least one compact divisor \footnote{Those numbers are not really the usual intersection number: usually one consider the restriction of a  Cartier divisor $D$ on a complete curve $C$ and then compute the degree of the corresponding line bundle on $C$ which is denoted as $D\cdot C$. In our case, $C$ is defined as the intersection of two divisors $D_i\cdot D_j$ so we have a triple intersection number. We do not have above interpretation of triple intersection number involving three non-compact divisor, and these numbers are coming from the consistent condition on the divisor relations.}.

\textbf{Mori cone}: Mori cone is an important structure of a algebraic variety. It is described in the toric case as the following simple set: 
\begin{equation}
\sum R_{>0} [V(\tau)],
\end{equation}
here $V(\tau)$ is the complete curve corresponding to a wall of a crepant resolution  $P^{'}$.  These walls are represented by
the internal edges of  $P^{'}$. The generators of  Mori cone can be found as follows.
For each internal edge $C_i$ which is a diagonal edge of a quadrilateral, we have the following relations for the four generators corresponding to the vertices of quadrilateral:
\begin{equation}
u_1^{(i)}+a u_2^{(i)} +b u_3^{(i)} + u_4^{(i)}=0.
\end{equation}
Not all of these relations are independent. Now a curve is the generator of Mori cone if the corresponding relation can not be written as the sum of other relations with \textbf{positive} coefficients. There are a total of $N+I-3$ generators, and the Mori cone is just $R_{N+I-3}^{+}$ once we identify those generators.

\textbf{Nef cone}: A divisor  $D=\sum a_i D_i$ is called Nef if it satisfies the following condition
\begin{equation}
D\cdot C\geq 0,
\end{equation}
for all irreducible complete curve $C$. The space of Nef divisors form a cone and called Nef cone. This cone is dual to the Mori cone, and is defined as 
\begin{equation}
(\sum a_i D_i) \cdot C_k \geq 0.
\end{equation}
Here $D_i$ is the basis of the divisors, $C_k$ is the generator of Mori cone. 
So each generator of Mori cone defines a hypersurface in the space of divisors, and this hypersurface gives a co-dimension one surface of the Nef cone.

\textbf{Prepotential}: Each crepant resolution and its Nef cone gives 
a chamber of Coulomb branch, and  the prepotential in this chamber is given by
\begin{equation}
{\cal F}={1\over 6}{\vec{D}}\cdot{\vec{D}}\cdot{\vec{D}}={1\over 6} (\sum D_i \phi_i)^3.
\end{equation}
Here $\vec{D}$ is a general point in Nef cone and $D_i$ denote the basis of the divisors. To compare our result  with the formula from the gauge theory, we need to use following expression from $SU(N)$ gauge theory. 
Let's denote $a_j=\phi_j-\phi_{j-1}$ with $\phi_N=\phi_{0}=0$. Consider $SU(N)$ SYM theory with CS term $k$ and $n_f$ fundamental flavors, we have (assume gauge coupling is zero):
\begin{align}
&{\cal F}={1\over6}[\sum_{i<j}(a_i-a_j)^3+k\sum_{i=1}^N a_i^3-{n_f\over2}\sum_{i=1}^N|a_i|^3]    \nonumber\\
&={1\over6}(8\sum_{j=1}^{N-1}\phi_j^3-3\sum_{j=1}^{N-1}(\phi_j^2\phi_{j+1}+\phi_j\phi_{j+1}^2)+3\sum_{j=1}^{N-1}(k+N-2j-1)(\phi_j^2\phi_{j+1}-\phi_j\phi_{j+1}^2)-{n_f\over2}\sum_{i=1}^N|a_i|^3]).
\end{align}

\textbf{Flop}: Different crepant resolutions are related by flops, see figure. \ref{flop}. Now consider a quadrilateral and the corresponding flop, we have two relations for two curves:
\begin{align}
&C:~~u_1-u_2-u_3+u_4=0 \nonumber \\
&C^{'}:~~-u_1+u_2+u_3-u_4=0 
\end{align} 
The constraint on the dual Nef cone is 
\begin{align}
&{\sum a_i D_i} \cdot C \geq 0 \rightarrow a_1-a_2-a_3+a_4 \geq 0, \nonumber\\
&{\sum a_i D_i} \cdot C^{'} \geq 0 \rightarrow -a_1+a_2+a_3-a_4 \geq 0.
\end{align}
 So they define the same hypersurface in the space of divisors, but the two chamber is living on different sides of this hypersurface. In fact, this hypersurface is shared by two Nef cones corresponding to these two crepant resolutions related by above flop!

\begin{figure}[h]
    \centering
    \includegraphics[width=2.5in]{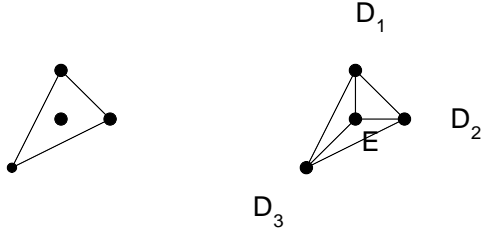}
    \caption{One toric diagram and its crepant resolution.}
    \label{example1}
\end{figure}

\textbf{Example 1}: Consider the toric diagram and its unique triangulation shown in figure. \ref{example1}.
The divisor class group is generated by torus invariant divisors $E, D_1, D_2, D_3$ associated with the ray generators of the fan. These divisors 
satisfy the relations:
\begin{equation}
E+D_1+D_2+D_3=0,~~D_2-D_3=0,~~D_1-D_3=0,
\end{equation}
and we have:
\begin{equation}
E=-3D_1,~~D_2=D_3=D_1.
\end{equation}
The triple  intersection numbers are
\begin{equation}
ED_1^2=1,~~E^2D_1=-3,~~E^3=9.
\end{equation}

There are three complete curve $C_i=E\cdot D_i$, and they are equal due to the divisor relations, i.e. $C_1=C_2=C_3=C$. The Mori cone is generated by  $C$ and is just the positive real line.
The Nef cone is computed as follows
\begin{equation}
a D_1 \cdot C= a \geq 0,\rightarrow a\geq 0.
\end{equation}
Now for a point in  Nef cone, the prepotential is 
\begin{equation}
{\cal F}={1\over 6}(\phi E)^3={3\over 2}(\phi^3),~~~-a=3\phi.
\end{equation}
The effective coupling constant is 
\begin{equation}
{1\over g^2}={1\over 2} {\partial^2 {\cal F}\over \partial^2 \phi}={9\over 2}\phi.
\end{equation}

\begin{figure}[h]
    \centering
    \includegraphics[width=3.0in]{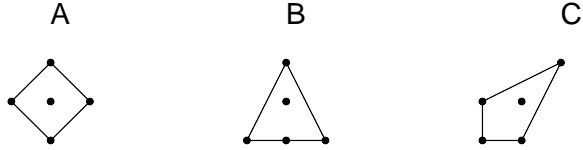}
    \caption{Three toric diagrams which have pure SU(2) gauge theory description along certain locus of Coulomb branch.}
    \label{example2}
\end{figure}

\textbf{Example 2}: We will compute the Coulomb branch of  three toric diagrams listed in figure. \ref{example2}.

\begin{figure}[h]
    \centering
    \includegraphics[width=4.0in]{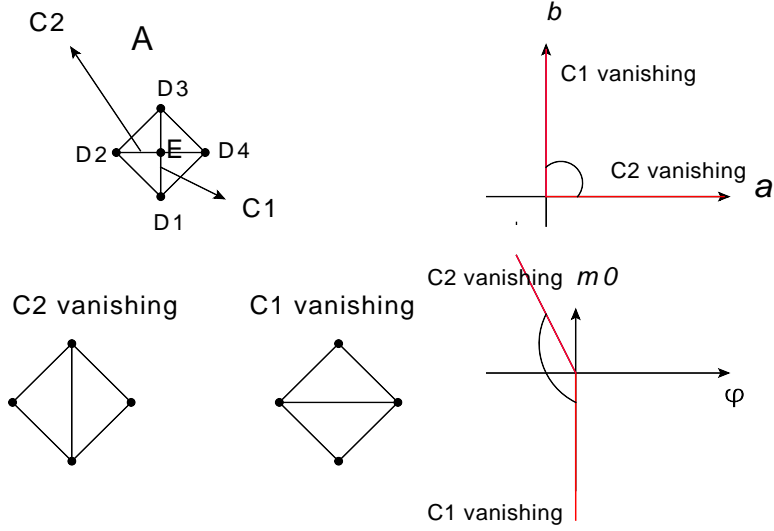}
    \caption{Left: The crepant resolution of toric diagram A in figure. \ref{example2} and two partial resolutions where we have a gauge theory description. 
    Right: Up is the Nef cone with basis of divisors $D_1$ and $D_2$; Lower is the cone using the basis $E$ and $-D_1$. }
    \label{puresu2}
\end{figure}

\textbf{A}:~~Consider  toric diagram A in  figure. \ref{example2} and its unique crepant resolution shown in figure. \ref{puresu2}. The relations between the divisors are:
\begin{equation}
E=-2D_1-2D_2,~~D_1=D_3,~~D_2=D_4.
\end{equation}
The triple intersection numbers are $E^3=8,~~E^2D_1= E^2 D_2= -2.$.

The complete curves are $C_1=E\cdot D_1,~C_2=E\cdot D_2,~C_3=E\cdot D_3,~C_4=E\cdot D_4$. They have the relations $C_1=C_3$ and $C_2=C_4$. So 
the Mori cone is generated by $C_1$ and $C_2$. 
Let's take $D_2$ and $D_1$ as the basis of divisor class. 
 The Nef cone is defined as 
\begin{align}
&(a[D_1]+b[D_2]) \cdot C_1\geq 0 \rightarrow a\geq 0,  \nonumber\\
&~(a[D_1]+b[D_2]) \cdot C_2\geq 0\rightarrow b\geq0 .
\end{align}
The prepotential is then
\begin{align}
&{\cal F}={1\over6}(a D_1+bD_2)^3={1\over 6}(E\phi-D_1m_0)^3={1\over6}(8\phi^3+6m_0\phi^2),~~\nonumber\\
&-a=2\phi+m_0\leq 0,~~-b=2\phi\leq0.
\end{align}
We change the coordinate to make the contact with the gauge theory result. 
The effective gauge coupling is 
\begin{equation}
g={1\over2}{\partial^2 {\cal F}\over \partial ^2\phi}=4\phi+m_0.
\end{equation}

\textbf{Remark 1}:  It is interesting to note that the range of $m_0$. One $SU(2)$ gauge theory 
description is valid in the range $\phi \leq 0$ and $m_0\leq 0$ (We choose a different Weyl chamber from the usual one where $m_0$ and $\phi$ are both non-negative.).  
In the other region $m_0>0$ we have a different gauge theory theory description.

\textbf{Remark 2}: The boundary of the Coulomb branch corresponds to the locus where there is extra massless particle. In our case, we know that it corresponds to 
shrink one of the cycle $C_1$ or $C_2$, and we get a pure $SU(2)$ gauge theory description.  We interpret the boundary defined by the equation $2\phi$ ($a=0$ and $C_1$ is vanishing) as the indication 
that a W boson become massless, and the boundary defined by the equation $2\phi+m_0=0$ ($b=0$ and $C_2$ is vanishing ) indicates that a instanton particle carry 2 electric charge and one instanton
charge become massless.

\begin{figure}[h]
    \centering
    \includegraphics[width=4.5in]{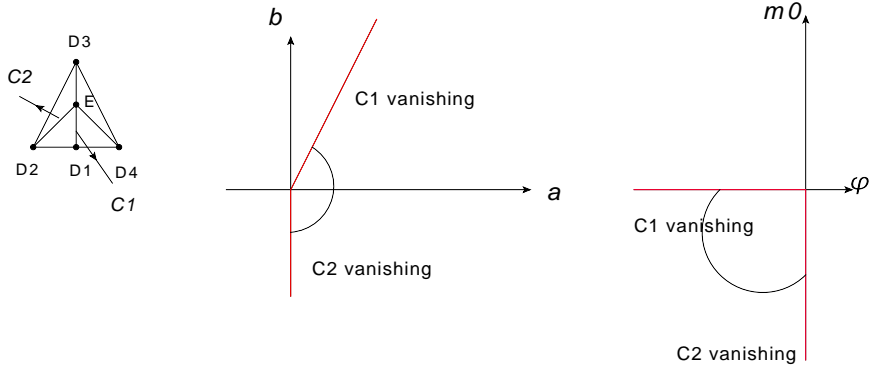}
    \caption{The crepant resolution of a toric diagram B in figure. \ref{example2} and its Coulomb branch.}
    \label{toriB}
\end{figure}

\textbf{B}:~~Next let's consider  toric diagram B in figure. \ref{example2} and its crepant resolution shown in figure. \ref{toriB}. We have the relation for the divisors:
\begin{equation}
E=-2D_1-4D_2,~~D_3=D_1+2D_2,~~D_4=D_2.
\end{equation}

The basis of the divisors are $D_1$ and $D_2$. We have the triple intersection number $E^3=8,~~ED_1^2=-2,~~E^2D_2=-2,$.
The complete curves are:
\begin{equation}
C_1=E\cdot D_1,~~C_2=E\cdot D_2,~~~~C_3=E\cdot D_3=C_1+2C_2.
\end{equation}
The Mori cone is generated by $C_1$ and $C_2$. The Nef cone is defined as
\begin{align}
&(a[D_1]+b[D_2])\cdot C_1=-2a+b\geq 0, \nonumber\\
&(a[D_1]+b[D_2])\cdot C_2=a\geq0. \nonumber\\
\end{align}
The prepotential can be written as 
\begin{equation}
{\cal F}={1\over6}(a[D_1]+b[D_2])^3={1\over6}(\phi E-m_0 D_2)^3={1\over6}(8\phi^{3}+6m_0\phi^{2}).
\end{equation}
Here $-a=2\phi,~~-b=4\phi+m_0$. The range of coordinates is 
\begin{equation}
\phi\leq 0,~~m_0\leq 0.
\end{equation}
The coupling constant is $g={1\over2}{\partial^2 {\cal F}\over \partial ^2\phi}=4\phi+m_0$.

\textbf{Remark}:  The prepotential of this example takes the exact same form as the above case. However, there are two important
differences. First, the range of the Coulomb branch is different. Second, we still have a massless W boson at the boundary $2\phi=0$ ($a=0$ with $C_2$ vanishing), but on 
the other boundary we have a particle with quantum number $m_0$ ($2a-b=0$ with $C_1$ vanishing) which carries no electric charge and one instanton charge to become massless. We 
do not see a $SU(2)$ gauge theory description on the other boundary. In the literature \cite{Taki:2014pba}, it is argued that this SCFT and the theory studied 
in last case describe the same SCFT. Our analysis of Coulomb branch  indicates that they are  different theories.

\begin{figure}[h]
    \centering
    \includegraphics[width=4.0in]{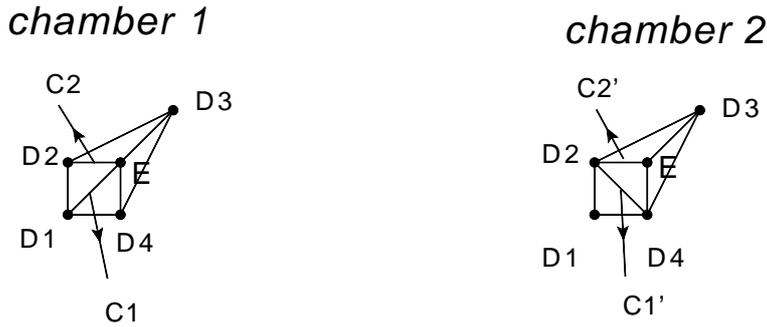}
    \caption{Two resolutions of toric diagram C in figure. \ref{example2}.}
    \label{su2c}
\end{figure}

\textbf{C}: Next let's consider toric diagram C shown in figure. \ref{example2} and its crepant resolution shown in left of figure. \ref{su2c}. We have the following relation of the divisors
\begin{equation}
E=-2D_1-3D_2,~~D_4=D_2,~~D_3=D_1+D_2.
\end{equation}
The basis of the divisor can be taken as $D_1$ and $D_2$. The triple intersection numbers are $E^3=8,~~E^2D_2=-2,~~E^2D_1=-1,~~ED_1^2=-1$.
The complete curves are 
\begin{equation}
C_1=E\cdot D_1,~~~C_2=E\cdot D_2,~~C_3=E\cdot D_3=C_1+C_2,~~C_4=E\cdot D_4=C_2.
\end{equation}
$C_1$ and $C_2$ are the generators for the Mori cone. The Nef cone is defined as:
\begin{align}
&(a[D_1]+b[D_2])\cdot C_1=-a+b\geq0,   \nonumber\\
&(a[D_1]+b[D_2])\cdot C_2=a\geq0.
\end{align}
The prepotential is 
\begin{align}
&{\cal F}={1\over 6}(aD_1+bD_2)^3={1\over6}(E\phi-D_2m_0)^3={1\over6}(8\phi^{3}+6m_0\phi^{2}), \nonumber\\
&-a=2\phi,~-b=3\phi+m_0.
\end{align}
For this chamber at the boundary ($a=0$ or $\phi=0$), we have $C_2$ vanishing, and we have a $SU(2)$ gauge theory description. At other boundary defined by $-a+b=0$, we have another particle with BPS mass formula $\phi+m_0$ to become massless. This is clearly different from the case $A$ and $B$.

\begin{figure}[h]
    \centering
    \includegraphics[width=4.0in]{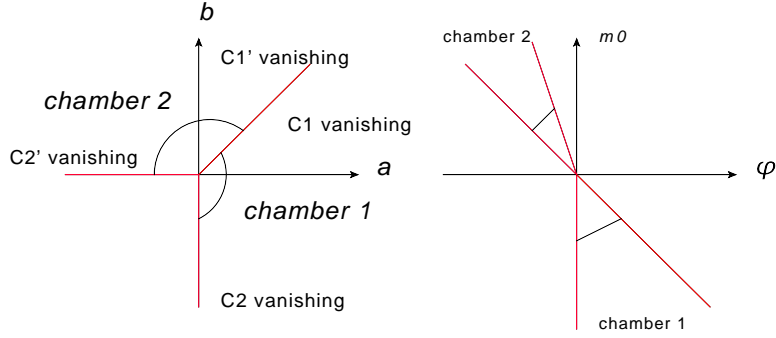}
    \caption{The Nef fan for the toric diagram C in figure. \ref{example2}, see figure. \ref{su2c} for the resolutions and notation for the curves.}
    \label{coulombC}
\end{figure}

For the other resolution shown in the right of figure. \ref{su2c}, the triple intersection numbers are $E^3=9,~~~E^2D_2=-3,~~ED_2^2=1$.
We have complete curves
\begin{equation}
C_1^{'}=D_4\cdot D_2,~~C_2^{'}=E\cdot D_2,~C_3^{'}=E\cdot D_3=C_2^{'},~C_4^{'}=E\cdot D_4=C_2^{'}.
\end{equation}
The Nef-cone is defined as:
\begin{align}
&(a[D_1]+b[D_2])\cdot C_1^{'}=a-b\geq0,   \nonumber\\
&(a[D_1]+b[D_2])\cdot C_2^{'}=b\ge0. \nonumber\\
\end{align}
The prepotential in this chamber is  
\begin{align}
&{\cal F}={1\over 6}(aD_1+bD_2)^3={1\over 6}(E\phi-D_2m_0)^3={1\over6}(9\phi^{3}+9m_0\phi^{2}+3m_0^2\phi)\nonumber\\
& -a=2\phi,~-b=3\phi+m_0
\end{align}
One boundary of this chamber is defined as $a-b=0$ which gives the same BPS particle from the first chamber. There is another boundary defined by $b=0$ (corresponds to $C_2^{'}$ vanishing). The full Coulomb branch is shown in figure. \ref{coulombC}.

\begin{figure}[h]
    \centering
    \includegraphics[width=6.0in]{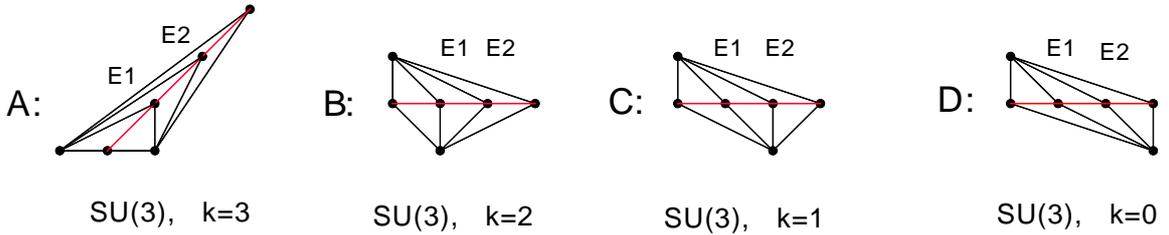}
    \caption{Four toric diagrams with their partial crepant resolutions. We also write down the gauge theory descriptions. }
    \label{su3}
\end{figure}

\textbf{Example 3}: 
Consider four toric diagrams and  their resolutions shown in figure. \ref{su3}, and let's consider only compact divisors. The prepotential is 
\begin{equation} 
{\cal F}={1\over6}(\phi_1^3 E_1^3+3\phi_1^2\phi_2 E_1^2E_2+3\phi_1\phi_2^2 E_1E_2^2+\phi_2^3E_2^3).
\label{pretoric}
\end{equation}
The data for intersection  numbers is shown in  table. \ref{table:su3}:
\begin{table}[h]
\begin{center}
\begin{tabular}{|c|c|c|c|c|}
\hline
  ~&$E_1^3$&$E_2^3$&$E_1^2E_2$&$E_1E_2^2$ \\ \hline
  A&8&8&-4&2 \\ \hline
  B&8&8&-3&1 \\ \hline
  C&8&8&-2&0 \\ \hline
  D&8&8&-1&-1 \\ \hline
\end{tabular}
\end{center}
\caption{The triple intersection number of the compact divisors for the toric diagrams in figure . \ref{su3}.}
  \label{table:su3}
\end{table}

For the prepotential for $SU(3)$ gauge theory with $n_f=0$ and CS level $k$, we get 
\begin{equation}
{\cal F}=8\phi_1^3+8\phi_2^3+3(k-1)\phi_1^2\phi_2+3(-k-1)\phi_1\phi_2^2.
\end{equation}
If $n_f=0$, the prepotential is invariant under the change $k\rightarrow -k, \phi_1\rightarrow \phi_2$. Comparing above formula
and the formula \ref{pretoric}, we can write down the CS level for the corresponding toric diagrams in figure. \ref{su3}.
We leave the detailed study of the Coulomb branch to the interested reader.

\begin{figure}[h]
    \centering
    \includegraphics[width=4.0in]{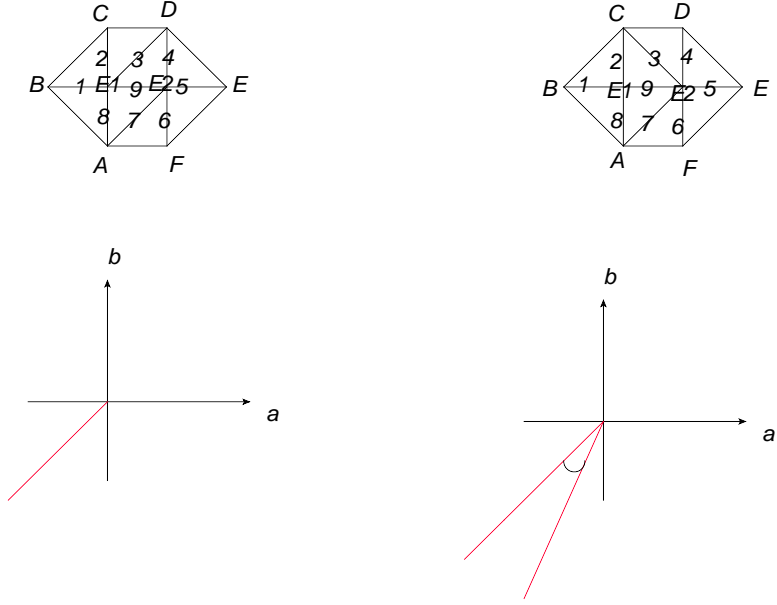}
    \caption{Two resolutions of a toric diagram and its Nef-cone.}
    \label{quiver}
\end{figure}

\textbf{Example 4}: Let's consider the toric diagram and one of its crepant resolution shown in left of figure. \ref{quiver}.  
The relations \footnote{Here the notation of divisors indicate their coordinates, one should not confuse it with the relations between divisors.} for the complete curves are 
\begin{align}
& ~~~C_1:~-2E_1+A+C=0,~~~~~~~~~~~~C_5:~-2E_2+D+F=0,        \nonumber\\
& \bullet C_2:~-C-E_1+B+D=0,~~~~\bullet C_6:~A+E-E_2-F=0,        \nonumber\\
& \bullet C_3:~E_2+C-D-E_1=0,~~~~~~~\bullet C_7:~E_1+F-A-E_2=0,         \nonumber\\
& ~~~C_4:~-2E_2+E_1+E=0,~~~~~~~~C_8:~-2E_1+B+E_2=0,         \nonumber\\
& \bullet C_9:~A+D-E_1-E_2=0.         
\end{align}

We take $C_2,C_3,C_6,C_7,C_9$ as the generator for the Mori cones, and other curves are expressed as:
\begin{equation}
C_1=C_3+C_9,~~C_4=C_6+C_7,~~C_5=C_7+C_9,~~C_8=C_2+C_3.
\end{equation}
We get a $SU(2)-SU(2)$ quiver gauge theory if $C_3,~C_7,~C_9$ are vanishing.  Now let's compute 
the Nef cone (let's take the mass parameters to be zero.), we have 
\begin{equation}
(aE_1+bE_2)\cdot C_i \geq 0,~~i=2,3,6,7,9
\end{equation}
and we get the constraints:
\begin{equation}
C_2:-a\geq 0,~C_3:-a+b\geq 0,~C_6:-b\geq0,~C_7:a-b\geq 0,~C_9:-a-b\geq 0.
\end{equation}
So this chamber is defined by the equations $a=b$ and $a\leq 0,~b\leq 0$. It is interesting to note that  
not all of the Coulomb branch of the $SU(2)-SU(2)$ gauge theory can be probed! The prepotential in this chamber is 
\begin{equation}
{\cal F}={1\over 6}(aE_1+bE_2)^3={1\over 6}(7a^3+7b^3-3a^2b-3ab^2).
\end{equation}

Next let's compute the Mori cone and Nef cone for the resolution shown on the right of figure. \ref{quiver}. The relation 
for the complete curves are
\begin{align}
& ~~~C_1:~-2E_1+A+C=0,~~~~~~~~~~C_5:~-2E_2+D+F=0,        \nonumber\\
& \bullet C_2:~--2E_1+B+E_2=0,~~~~\bullet C_6:~A+E-E_2-F=0,        \nonumber\\
& \bullet C_3:~-E_2-C+D+E_1=0,~~\bullet C_7:~E_1+F-A-E_2=0,         \nonumber\\
& ~~~C_4:~-2E_2+E_1+E=0,~~~~~~~C_8:~-2E_1+B+E_2=0,         \nonumber\\
& \bullet C_9:~-2E-1+A+C=0.         
\end{align}
The generators for the Mori cone are $C_2,C_3,C_6,C_7,C_9$, and the other curves are expressed as 
\begin{equation}
C_1=C_9,~~C_8=C_2,~~C_4=C_6+C_7,~~C_5=C_3+C_9.
\end{equation}
One also has the $SU(2)-SU(2)$ quiver gauge theory description when $C_3,~C_7,~C_9$ vanish. The Nef cone is again  defined as 
\begin{equation}
(aE_1+bE_2)\cdot C_i \geq 0,~~i=2,3,6,7,9
\end{equation}
and we have the constraints:
\begin{equation}
C_2:-2a+b\geq 0,~C_3:a-b\geq 0,~C_6:-b\geq0,~C_7:a-b\geq 0,~C_9:-2a\geq 0.
\end{equation}
The corresponding cone is shown in figure. \ref{quiver}, and one of the boundary corresponds to vanishing of $C_2$ (one find a $SU(2)$ gauge group here), while the other boundary corresponds to vanishing of $C_3$ and $C_7$ .  The prepotential in this chamber is 
\begin{equation}
{\cal F}={1\over 6}(aE_1+bE_2)^3=8a^3+6b^3-6a^2b.
\end{equation}

\textbf{Remark}: It was argued in \cite{Morrison:1996xf,Intriligator:1997pq} that the UV limit of a quiver gauge theory does not define a SCFT since the prepotential can not be convex in the whole Coulomb branch of gauge theory which is identified with the Weyl Chamber. Our computation above shows that actually the range of Coulomb branch is not the naive Weyl chamber of 
the gauge theory.

\subsubsection{Relation to $(p,q)$ web and gauge theory construction}
Given a crepant resolution of a toric  Gorenstein singularity, one can construct a dual diagram: put a trivalent vertex 
inside a triangle with edges perpendicular to the boundaries of triangle, and one get a $(p,q)$ web by connecting those trivalent vertices. So a $(p,q)$ web is completely equivalent to a toric Gorenstein singularity. 

As we see from above example, given a crepant resolution and it is possible to find a non-abelian gauge theory at 
certain sub-locus of the Coulomb branch. However, the SCFT is not completely characterized by the non-abelian gauge theory description  as SCFT has other massless particles from the boundary of the Coulomb branch which can not be determined by the gauge theory.

\subsubsection{Classification: lower rank theory}
For toric Gorenstein singularity, the classification is reduced to the classification of two dimensional convex lattice polygone up to following integral unimodular affine transformation on the plane:
\begin{equation}
\left(\begin{array}{c}
x^{'}\\
y^{'}\\
1
\end{array}\right)=
\left(\begin{array}{ccc}
a&b&e\\
c&d&f\\
0&0&1
\end{array}\right)\left(\begin{array}{c}
x\\
y\\
1
\end{array}\right)
\end{equation}
here $(a,b,c,d,e,f)$ are integers and $ad-bc=1$.  It is easy to check the number of boundary points, internal points and the area of the polygon is not changed under above transformation.
There are also some interesting facts about the convex lattice polygon, and here we collect two of the useful ones:
\begin{enumerate}
\item The area of a lattice polygon is given by the Pick's formula
\begin{equation}
A={B\over 2}+I-1.
\end{equation}
Here $B$ is the number of boundary points and $I$ is the number of interior points. 
\item The boundary points are constrained by the interior points, i.e.
\begin{equation}
B\leq 2I+6,~~~I>1,
\end{equation}
and $B\leq 2 I+7$ for $I=1$.
\end{enumerate}

\begin{figure}[h]
    \centering
    \includegraphics[width=4.0in]{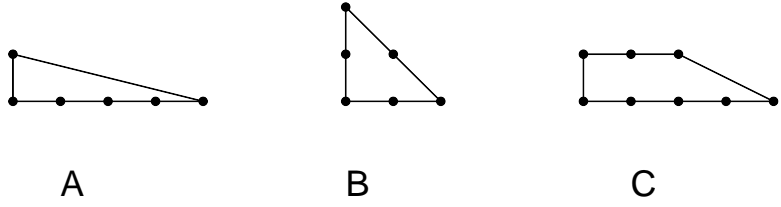}
    \caption{Toric diagram of rank zero theory.}
    \label{rank0}
\end{figure}

\textbf{Rank zero theory}
The rank zero theory is classified by the convex polygon without any interior point. They were classified in \cite{rabinowitz1989census}. By TRIANG(p,h) we denote the triangle whose vertices are $(0,0), (p,0)$, and $(0,h)$. By TRAP(p,q,h) we denote the 
trapezoid whose vertices are $(0,0), (p,0), (0,h)$ and $(q,h)$. Then we have the following classification: if K is a convex lattice polygon with $g=0$ (g denotes the number of interior point), then K is lattice equivalent to one of the following polygons
\begin{enumerate}
\item TRIANG(p,1), where $p$ is any positive integer. The flavor symmetry is $SU(p)$. We interpret this theory as ``zero" flavor of $SU(p)$ group.
\item TRIANG(2,2). The flavor symmetry is $SU(2)\times SU(2)\times SU(2)$. We interpret this theory as the tri-fundamental of SU(2) groups.
\item TRAP(p,q,1), where $p$ and $q$ are any positive integers. The flavor symmetry is $SU(p)\times SU(q)\times U(1)$. This is the bi-fundamental for $SU(p)\times SU(q)$. 
\end{enumerate}
See figure. \ref{rank0} for some examples.
With the field theory interpretation of these rank one zero toric diagram, one can easily write down the gauge theory descriptions from the partial resolution of the general toric Gorenstein singularity. 

\begin{figure}[h]
    \centering
    \includegraphics[width=4.5in]{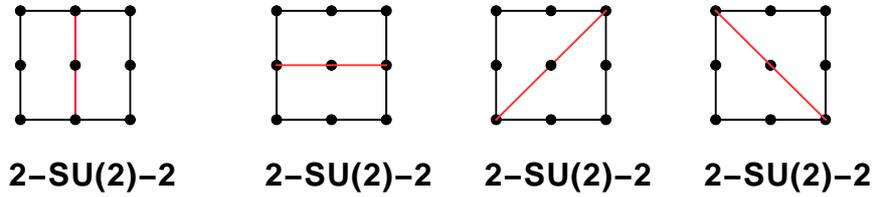}
    \caption{Non-abelian gauge theory descriptions for a rank one toric diagram.}
    \label{gauge}
\end{figure}

\textbf{Example}: See figure. \ref{gauge} for an example, and we list four gauge theory descriptions.

\begin{figure}[h]
    \centering
    \includegraphics[width=6.0in]{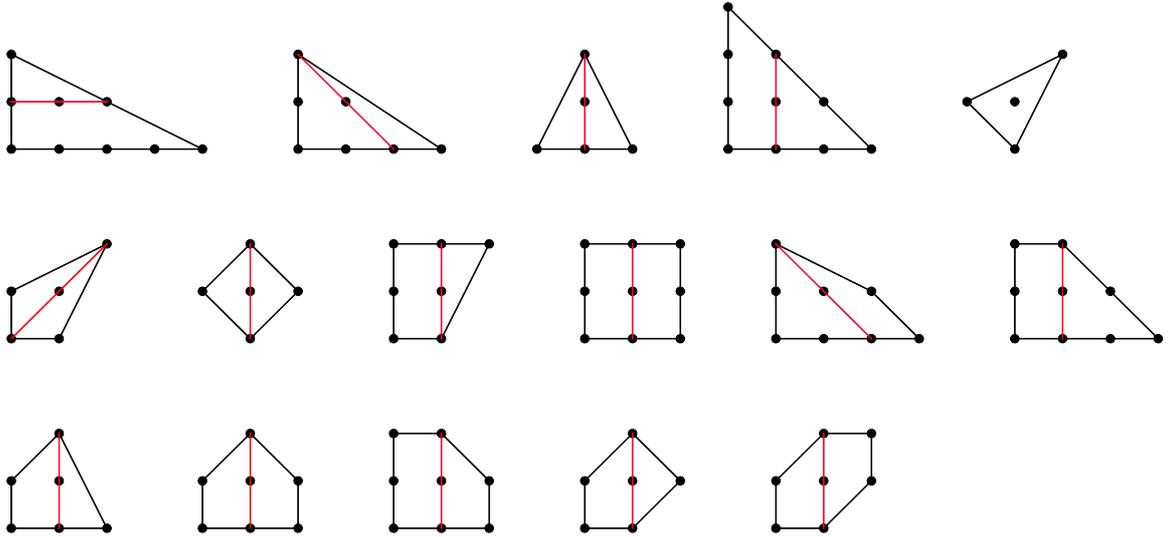}
    \caption{Toric diagram of rank one theory. The red lines indicate the partial resolution which give a gauge theory description.}
    \label{rank1}
\end{figure}

\textbf{Rank one theory}
The rank one theory is classified by the convex polygon with only one interior point, see figure. \ref{rank1}. We also show the 
partial resolution and the interested reader can write down the gauge theory description. 

\begin{figure}[h]
    \centering
    \includegraphics[width=5.0in]{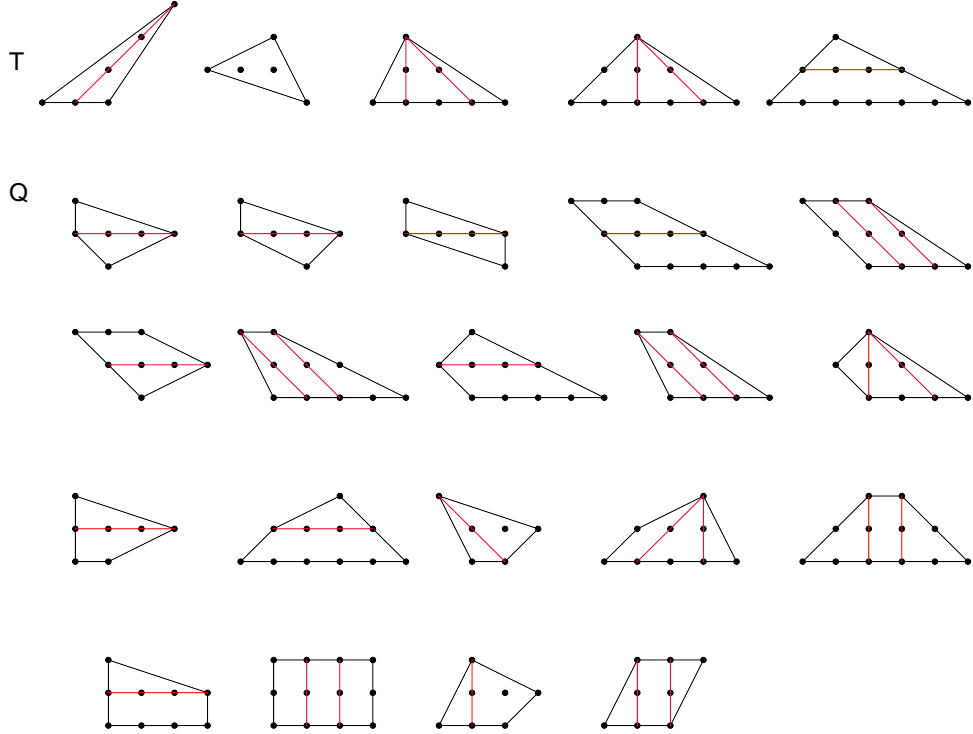}
    \caption{Part A of toric diagram of rank two theory. The red lines indicate the partial resolution which  give a gauge theory description.}
    \label{rank2A}
\end{figure}

\begin{figure}[h]
    \centering
    \includegraphics[width=5.0in]{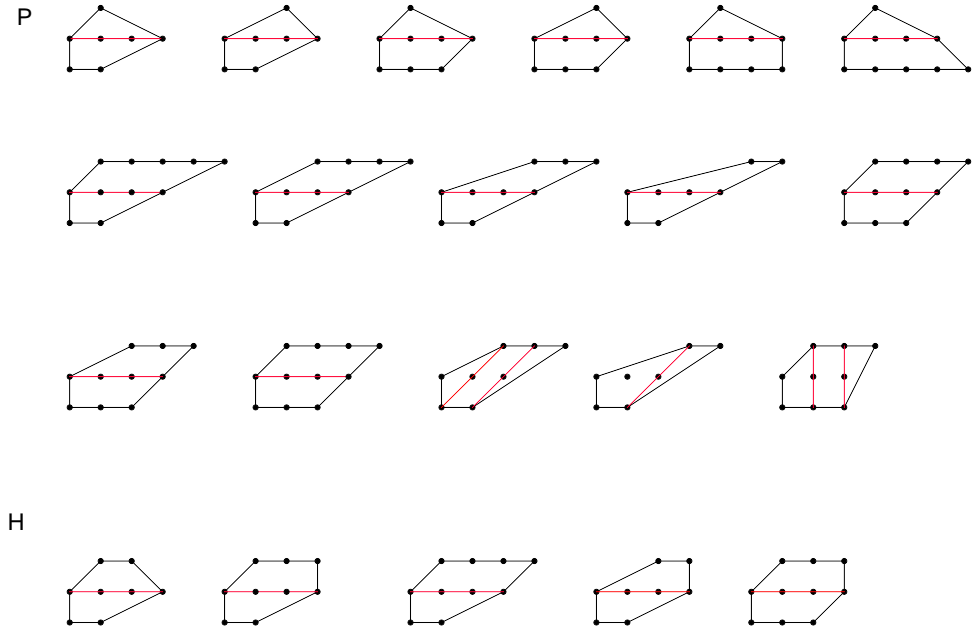}
    \caption{Part B of toric diagram of rank two theory. The red lines indicate the partial resolution which will give a gauge theory description.}
    \label{rank2B}
\end{figure}

\textbf{Rank two theory}
The rank two theory is classified by the convex polygon with two interior points \cite{wei2012lattice}, see figure. \ref{rank2A} and figure. \ref{rank2B}. We also write down one partial resolutions for them. 

\subsection{Toric Q-Gorenstein  singularity}
 One can also consider  Q-Gorenstein toric singularity, and the computation of Mori and Nef cones are similar. The major interesting 
 new feature is that the generic point of Coulomb branch has an interacting part described by the Q-factorial terminal toric singularity. It would be interesting to further study the theories defined by those singularities.

\section{Quotient singularity}
Another important class of 3-fold canonical singularity is quotient singularity $C^3/G$ where $G$ is a finite subgroup of $SL(3,C)$. 
All such finite subgroups have been listed in \cite{yau1993gorenstein}. If $G$ is abelian, such singularity is also toric, and we can use the toric method to 
study them. For more general class of singularities,
We would like to point out some important properties associated with such singularities:
\begin{itemize}

\item \textbf{Flavor symmetry}: Let $G\in GL(n,C)$ be a small subgroup, let $S=\{z\in C^3: g(z)=z\}$. Then the singular locus of $V_G$ is $S/G$.  Since 
we are interested in finite subgroup of $SL(3,C)$, the singular locus is at least co-dimension two. The singularity over 
such one dimensional singular locus is two dimensional $ADE$ singularity, and they give  the corresponding non-abelian flavor symmetries. 

In fact, three dimensional quotient singularity is isolated if and only if $G$ is an abelian subgroup and $1$ is not an eigenvalue of $g$
for every nontrivial element of $g$ in $G$.

\item \textbf{Crepant resolution}: There exists  crepant resolution for the quotient singularity \cite{roan1996minimal}, and the number of crepant divisors are
related to the representation theory of  finite group $G$.

\end{itemize}

\section{Hypersurface singularity}
A third class of 3-fold canonical singularity can be defined by a single equation and we call them hypersurface singularity. 
Let's consider a hypersurface singularity $f: C^4\rightarrow C$, and further impose the condition that $f$ is isolated, i.e. 
equations $f=0$ and ${\partial f\over \partial z_i}=0, i=0,\ldots, 3$ have a unique solution at the origin. We also impose the condition that 
$f$ has a $C^*$ action
\begin{equation}
f(\lambda^{q_i} z_i)=f(z_i),~~~q_i>0.
\end{equation}
The rational condition imposes the constraint on $q_i$:
\begin{equation}
\sum q_i >1.
\end{equation}
Since hypersurface singularity is Gorenstein, so the above rational condition also implies that it is canonical! The partial crepant resolution 
for these singularities can be found using the blow-up method. We'd like to compute the local divisor class group and the number of crepant divisors. 

We could consider more general isolated hypersurface singularity. 
Let's consider an isolated hypersurface singularity $f: C^{n+1}\rightarrow C$. Isolated singularity implies that 
equations $f=0$ and ${\partial f\over \partial z_i}=0, i=1,\ldots, n+1$ have a unique solution at the origin.  We may represent such an isolated singularity by 
a polynomial
\begin{equation}
f=\sum_{\nu \in N^{n+1}} a_{\nu} z^{\nu}.
\end{equation}
We set
\begin{align}
& \text{supp}~f=\{\nu \in N^{n+1}| a_\nu \neq 0 \} \nonumber\\
& \Gamma_{+}(f): \text {convex hull of}~\cup _{\nu \in \text{supp}f} (\nu+R_{+}^{n+1}) \nonumber\\
&\Gamma(f): \text{union of compact faces of}~\Gamma_{+}(f)
\end{align}
Let $\sigma$ be a face of $\Gamma(f)$, set $f_\sigma=\sum_{\nu \in \sigma} a_{\nu} x^{\nu}$.  A polynomial $f$ is called \textbf{Newton non-degenerate} if the critical points of $f_{\sigma}$ do not consist of points 
in ${C^{*}}^{n+1}$ (points  with all coordinates non-zero) for all the faces in $\Gamma(f)$. We may further assume that $f$ is \textbf{convenient} in the sense that its 
support intersects with the coordinate axis, and this assumption does not change any generality. 

Given a \textbf{Newton non-degenerate} singularity $f$, one can define a Newton order on the polynomial ring $C[z_1,\ldots, z_{n+1}]$. For $i$th $n$ dimensional face $\sigma_i$ of
$\Gamma(f)$, one can define a $C^*$ action $l_i$ such that 
$l_i(f_{\sigma_i})=1$. For a monomial $\nu=\sum_{j=1}^{n+1}z_j^{m_j}$, we define its weight  as
\begin{equation}
\alpha(\nu)=min_{i}(l_i(\nu)).
\end{equation}
We notice that  $w(\nu)=l_i(\nu)$ if $\nu$ is in cone $(0,\sigma_i)$.  For a polynomial $g$ in $C[z_1,\ldots, z_{n+1}]$, we define 
\begin{equation}
\alpha(g)=min\{w(\nu)|\nu \in \text{supp}~g\}.
\label{newtonorder}
\end{equation} 

For such an singularity, we can define two algebras  called Milnor algebra  $J_f$ and Tjurina algebra $T_f$:
\begin{align}
&J_f={C[z_1,\ldots, z_{n+1}]\over \{ {\partial f\over \partial z_1}, \ldots,{\partial f\over \partial z_{n+1}}\}}, \nonumber\\
&T_f={C[z_1,\ldots, z_{n+1}]\over \{f,{\partial f\over \partial z_1}, \ldots,{\partial f\over \partial z_{n+1}}\}}.
\end{align}
These two algebras are finite dimensional since $f$ defines an isolated singularity. $\mu=dim(J_f)$ is called Milnor number and $\tau=dim(T_f)$ is called Tjurina number. 
Obviously $\mu\geq \tau$, and $\mu=\tau$ if and only if the singularity is \textbf{quasi-homogeneous} (or semi-quasihomogeneous). 

\begin{center}
\begin{figure}[h]
\small
\centering
\includegraphics[width=5cm]{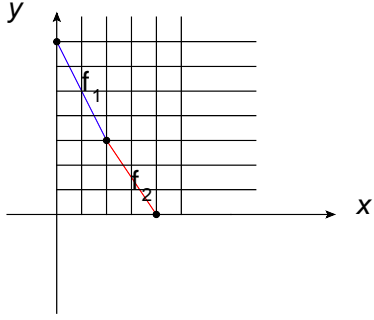}
\caption{Newton polyhedron for singularity $f=x^4+y^7+x^2y^3$.}
\label{newton}
\end{figure}
\end{center}

\textbf{Example}: Given an isolated singularity $f=x^4+y^7+x^2 y^3$, the newton polyheron is shown in figure. \ref{newton}.  
The Milnor number and Tjurina number are  $\mu=16, \tau= 14$. The $C^*$ actions from  $f_{1}$ and $f_{2}$ are $l_1(x)=1/4, l_1(y)=1/6$ and 
$l_1(x)=\frac{2}{7}, l_1(y)={1\over 7}$.  For a monomial $xy$, we have $l_1(xy)={3\over7},~l_2(xy)={5\over12}$, so $\alpha(xy)=min({3\over7},{5\over12})={5\over12}$. 

Let's focus on 3 dimensional singularity from now on, i.e. $n=3$.  Given a Newton non-degenerate singularity, one can define a singularity spectrum $S(f)$ using the Newton filtration defined above. For the quasi-homogeneous singularity, the singularity spectrum can be found easily. Let's take a monomial basis $\phi_{i}, i=1,\ldots,\mu$ of the Jacobian algebra $J_f$, the spectrum is given by the following formula
\begin{equation}
S(f)=\alpha(\phi_i)+\sum q_i -1,~~\phi_i\in J_f
\end{equation}
We can denote the spectrum $S(f)$ as an ordered set $(\alpha_1,\ldots, \alpha_\mu)$, and 
 rational condition implies $\alpha_1>0$, which implies $\sum q_i -1>0$ \footnote{Since $1$ is always the generator of $J_f$ with minimal weight, and we have  $\alpha_1=\sum q_i-1$}. 
 
 For general Newton non-degenerate singularity, it is possible also to find a regular monomial basis $B$ of $J_f$ such that the singularity spectrum is given by 
 \begin{equation}
 S(f)=\alpha(m+{\bf{1}})-1,~~m\in B
 \end{equation}
 Here $m$ denotes the exponent of the monomial basis, and $1$ denotes the vector $(1,1,1,1)$.  Rational condition implies that $\alpha_1>1$. The divisor class group $\rho(X)$ can be expressed \cite{caibuar2003divisor} as follows
\begin{equation}
\rho(X)= \{\# \alpha_i=1,~~\alpha_i\in S(f)\}.
\end{equation}
namely we count the number of ones in singularity spectrum $S(f)$.   

\textbf{Remark}: These 3d canonical quasi-homogeneous singularity has been used in \cite{Xie:2015rpa} to study four dimensional $\mathcal{N}=2$ SCFT. In that context, the weights with value $1$ in singularity spectrum give the mass parameters.

Let's define a crepant weighting as the weights $w({\bf{1}})=w(f)+1$ (here the weighting has positive integral weights on coordinates). We denote all such weighting as $W(f)$. 
The number of crepant divisors are found as \cite{caibuar2003divisor}:
\begin{equation}
c(X)=\sum_{\alpha \in W(f), dim \Gamma_\alpha=1} length \Gamma_{\alpha}+\#\{\alpha\in W(f): dim \Gamma_{\alpha}\geq 2\},
\end{equation}
where $\Gamma_\alpha$ denotes the face of $\Gamma(f)$ corresponding to $\alpha$.

More generally, one could consider isolated complete intersection canonical singularity which is defined by two polynomials $(f_1,f_2)$. Those complete intersection singularities with $C^*$ action
has been classified in \cite{Chen:2016bzh}. The divisor class group for such quasi-homogeneous singularity has been computed in \cite{flenner1981divisorenklassengruppen}.

\section{Conclusion}
We argue that every 3-fold canonical singularity defines a five dimensional $\mathcal{N}=1$ SCFT. We can read off
the non-abelian flavor symmetry from the singularity structure, and the Coulomb branch is described by the crepant resolutions of the singularity. For each crepant resolution, one can compute Nef cone which describes 
a chamber of Coulomb branch. The faces of  Nef cone correspond to the locus where an extra massless particle appears.
Different resolutions are related by flops, and these Nef cones form a fan which is claimed to be the full Coulomb branch. 

We have given a  description for the Coulomb branch of several simple theories defined by toric singularity. 
Detailed computations for prepotential and Coulomb branch structure  for other interesting toric theories will appear elsewhere. The non-toric example would be also interesting to study too, i.e. complex cone over Fano orbifolds should be  interesting to study. 

It would be interesting to study other properties of these theories such as superconformal index and Seiberg-Witten curve for 5d theory on circle.

\section*{Acknowledgements}
DX would like to thank H.C Kim, P. Jefferson, K. Yonekura for helpful discussions. 
DX would like to thank Korea Institute for Advanced Study, University of Cincinatti, Aspen center for physics for hospitality during various stages of this work. 
The work of S.T Yau is supported by  NSF grant  DMS-1159412, NSF grant PHY-
0937443, and NSF grant DMS-0804454.  
The work of DX is supported by Center for Mathematical Sciences and Applications at Harvard University, and in part by the Fundamental Laws Initiative of
the Center for the Fundamental Laws of Nature, Harvard University. 

\bibliographystyle{utphys}

\bibliography{5d}

\end{document}